\documentclass[useAMS,usenatbib,usegraphicx]{mn2e}

\def\msun{$\rm M_{\sun}$~}

\newcommand{\oiii}{[O$\:${\small III}]~}
\newcommand{\hbeta}{H$\beta$~}

\title[Molecular gas and star formation in the red-sequence
  counter-rotating disc galaxy NGC 4550]{Molecular gas and star formation in the red-sequence
  counter-rotating disc galaxy NGC 4550}

\author[A. F. Crocker et al.]{Alison F. Crocker,$^{1}$ Hyunjin Jeong,$^{2}$ Shinya
  Komugi,$^{3,4}$ Francoise Combes,$^{5}$ \newauthor Martin
  Bureau,$^{1}$  Lisa M. Young,$^{6}$ Sukyoung Yi$^{2}$ \\
$^{1}$Sub-Department of Astrophysics, University of Oxford, Denys
  Wilkinson Building, Keble Road, Oxford OX1 3RH \\
$^{2}$Department of Astronomy, Yonsei University, Seoul 120-749, Korea\\
$^{3}$Institute of Astronomy, University of Tokyo, 2-21-1 Osawa,
  Mitaka-shi, Tokyo 181-8588, Japan\\
$^{4}$National Astronomical Observatory of Japan, 2-21-1 Osawa,
  Mitaka-shi, Tokyo 181-8588, Japan\\
$^{5}$Observatoire de Paris, LERMA, 61 Av. de l'Observatoire, 75014,
Paris, France\\
$^{6}$Physics Department, New Mexico Institute of Mining and Technology, Socorro, NM 87801, U.S.A. }

\begin{document}

\date{}

\pagerange{\pageref{firstpage}--\pageref{lastpage}} \pubyear{}

\maketitle

\label{firstpage}

\begin{abstract}
We present observations of the CO(1-0) emission in the central 750~pc
(10\arcsec) of the counter-rotating disc galaxy NGC~4550, obtained at
the Institut de Radioastronomie Millim\'etrique (IRAM) Plateau de Bure
Interferometer. Very little molecular gas is detected, only
$1\times 10^{7}$ $\rm M_{\sun}$, and its distribution is lopsided,
with twice as much molecular gas observed at positive relative
velocities than at negative relative velocities. The velocity gradient
in the CO(1-0) emission shows that the molecular gas rotates like the
thicker of the two stellar discs, which is an unexpected alignment of
rotations if the thinner disc was formed by a major gas accretion event.
However, a simulation shows that the gas rotating like the thicker
disc naturally results from the coplanar merger of two
counter-rotating disc galaxies, demonstrating the feasibility of this
scenario for the formation of NGC~4550. We investigate various star
formation tracers to determine whether the molecular gas in NGC~4550
is currently forming stars. UV imaging data and optical absorption
linestrengths both suggest a recent star formation episode; the
best-fitting two population model to the UV-optical colours yields a
mass of young stars of $5.9 \times 10^{7}$ M$_{\sun}$ with an age of
280 Myr. The best 
information on the current star formation rate is a far infrared-based
upper limit of only 0.02 M$_{\sun}$ yr$^{-1}$. We are thus witnessing NGC~4550
either in a dip within a bursty star formation period or during a more
continuous low-level star formation episode.

  \end{abstract}


\section[]{Introduction}

Deciphering star formation histories of early-type galaxies is
important for understanding their formation and evolution.
While early-type galaxies were traditionally considered to
be simple single stellar populations, it is now clear that their star
formation histories must be more complicated. Both absorption
linestrengths \citep[e.g.][]{trager00} and UV-optical colours
\citep{yi05,kaviraj07} indicate that a significant portion  of
early-type galaxies have recently formed stars ($\approx30\%$
from the UV-optical colours). Reassuringly, molecular gas, the raw material for star formation, is
also found in early-type
galaxies \citep{henkel97}. Two recent surveys give detection rates of 28\% for
E/S0s in the SAURON sample \citep{combes07} and 78\% for S0s in a
volume-limited sample \citep{sage06}. 

NGC~4550 is an unusual galaxy (see Table 1), containing two coplanar
counter-rotating 
stellar exponential discs with nearly identical scale lengths \citep*{rubin92,rix92}. However, integral field unit data shows that one disc is
thicker and has a higher velocity dispersion than the other
\citep{kenney00, emsellem04}. Schwarzschild modelling by \citet{cappellari07} confirms the difference in scale heights and additionally shows that the two discs have equal mass within the SAURON field-of-view. The presence of these two discs with different scale
heights may cause the failure of some bulge-disc
decompositions. \citet{rix92} find a bulge to disc ratio of only 0.19
while \citet{scorza98} report a ratio of 5. The
S\'ersic fit of \citet{ferrarese06} gives a best-fit S\'ersic shape
parameter of 1.7, 
demonstrating that the surface brightness profile is not dominated by
a de Vaucouleurs bulge. With an effective
$|\rm B - \rm V|$ colour of 0.890 mag, NGC~4550 is very likely a red sequence
galaxy, which would make it one of a very small population of disc-dominated
red-sequence galaxies.  

Two main scenarios exist for explaining the counter-rotating stellar discs in
NGC~4550. \citet{rubin92} first suggested that accretion of
counter-rotating gas and subsequent star formation could create a
second disc rotating as observed. This gas accretion scenario has been
investigated by \citet{thakar98}, who find it difficult to
create extended, exponential gas discs in their
simulations. However, recent observations of NGC~5719
show an extended counter-rotating and star-forming gas disc clearly
accreted from external gas \citep{vergani07}, demonstrating the feasibility
of this origin. The second scenario invokes a 
coplanar major merger of two counter-rotating disc galaxies. While
requiring a precise alignment of the two discs to avoid excessive
heating, this merger scenario has been shown to produce resultant
galaxies resembling NGC~4550 \citep{puerari01}.

The ionised gas in NGC~4550 co-rotates with the thicker disc
\citep{rubin92, sarzi06}, not with the thinner, colder disc. An irregular dust
distribution is also observed 
within the central 20\arcsec~in diameter. The dust is distributed in
clumpy arcs and is stronger on the northern half of the galaxy
\citep{wiklind01}. H{\small{I}} observations have not detected neutral gas in
NGC~4550, the strictest upper limit being $7 \times 10^{7}$ \msun from
\citet{duprie96}. Molecular gas was first discovered in NGC~4550 by 
\citet{wiklind01} using the IRAM 30m telescope. They reported a small
mass of molecular gas, $1.3 \times 10^{7}$ $\rm M_{\sun}$, noting that the
distribution of the molecular gas is likely to resemble that of the dust,
due to a strong asymmetry in the observed CO emission line toward
positive relative velocities. 

Here we present interferometric observations of the CO emission in NGC~4550,
deriving the extent, kinematics and total mass of the molecular gas.
To investigate what the kinematics of the gas reveal
about the formation history of this unusual galaxy, we run a major merger
simulation (importantly including gas) and describe the features of its
evolution, comparing the results to observations. We also analyse various star formation indicators to
determine whether the molecular gas in NGC~4550 alerts us to another
early-type galaxy with ongoing star formation, or whether the
molecular gas instead appears to be stable. 

\begin{table}
 \caption{Basic properties of NGC~4550. The left and middle columns
 list the different quantities and their values; the right column
 lists the corresponding references.}
 \begin{tabular}{@{}llr} 
  \hline
  Quantity & Value & Ref. \\
  \hline
  R.A. (J2000.0) & 12$^{h}$ 35$^{m}$ 30.6$^{s}$ & 1\\
  Dec. (J2000.0) & +12$^{\circ}$ 13$\arcmin$ 15$\arcsec$ & 1\\
  Heliocentric velocity & $435$ km s$^{-1}$ & 2\\
  Distance & $15.5$ Mpc & 3\\
  Scale & $1\arcsec = 75$~pc & 3\\
  Type & SB0 & 4\\
  Corrected apparent B mag & $12.31$ & 5\\
  Corrected absolute B mag & $-18.64$ & 5\\
  (B-V)$_{e}$ & 0.89 & 5\\
  L$_{\mathrm{B}}$ & $4.3\times 10^{9}$ L$_{\sun}$ & 6\\
  L$_{\mathrm{FIR}}$ & $5.8\times10^{7}$ L$_{\sun}$ & 6\\
  L$_{\mathrm{FIR}}$/L$_{\mathrm{B}}$ & $1.4\times10^{-2}$ & 6\\
  L$_{\mathrm{FIR}}$/M$_{\mathrm{H}_{2}}$ & $8.0$
  L$_{\sun}$/M$_{\sun}$ & 6\\ 

  \hline
 \end{tabular}
\label{tab:basic}

References: (1) NED; (2) Derived from SAURON stellar kinematic data
\citep{emsellem04} (3) \citet{mei07};
 (4) \citet{devaucouleurs91}; (5) HyperLEDA;
(6) Derived quantity using data from NED, IRAS \citep{moshir90} and this paper.
\end{table}

\section[]{Plateau de Bure observations}
\subsection[]{Observations}

\begin{table}
\caption{CO observation calibrators}
\begin{tabular}{@{}lc}
\hline
 Type & Calibrators\\
\hline
 Bandpass & 3C273\\
 Phase & 3C273, 1156+295\\
 Flux & 3C273, 1156+295, 0528+134, 3C84 \\
\hline
\end{tabular}
\end{table}

\begin{figure} 
\begin{center}
  \rotatebox{270}{\includegraphics[width=6.5cm]{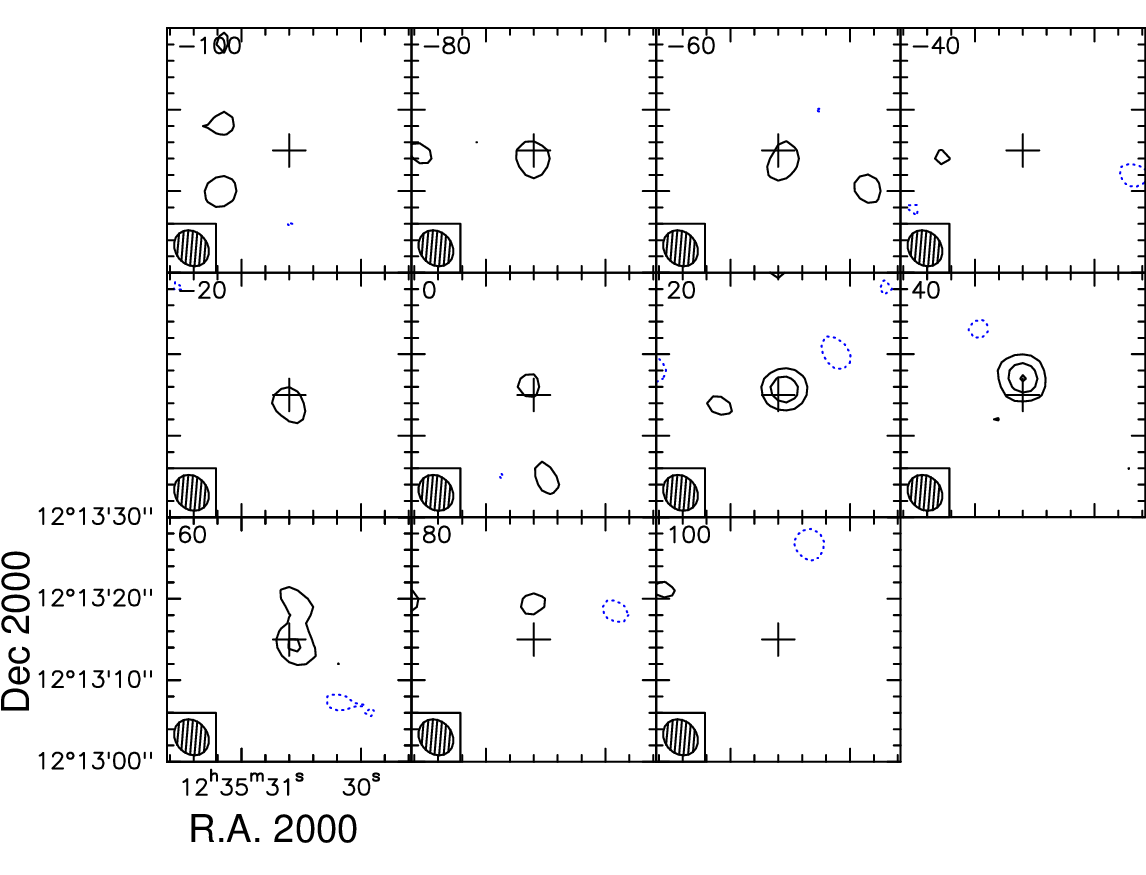}}
\caption{Channel maps of NGC~4550. The channels are
  20 km s$^{-1}$ wide and contours are plotted at -3 (dashed), 3, 6,
  and 9 $\sigma$ with $\sigma =$ 2.77 mJy beam$^{-1}$. The number in
  the top left corner of each frame is the central velocity of that
  frame, relative to the observed central velocity of
  NGC~4550 ($V_{\mathrm{sys}}=435$ km s$^{-1}$, determined using
  optical absorption lines). The synthesized beam is shown in the
  bottom left corner of each frame. The cross represents the centre of
  the galaxy as given by 2MASS.\label{fig:chmap10}} 
\end{center}
\end{figure}

 \begin{figure*}
\begin{center}
\rotatebox{270}{\includegraphics[width=7cm]
  {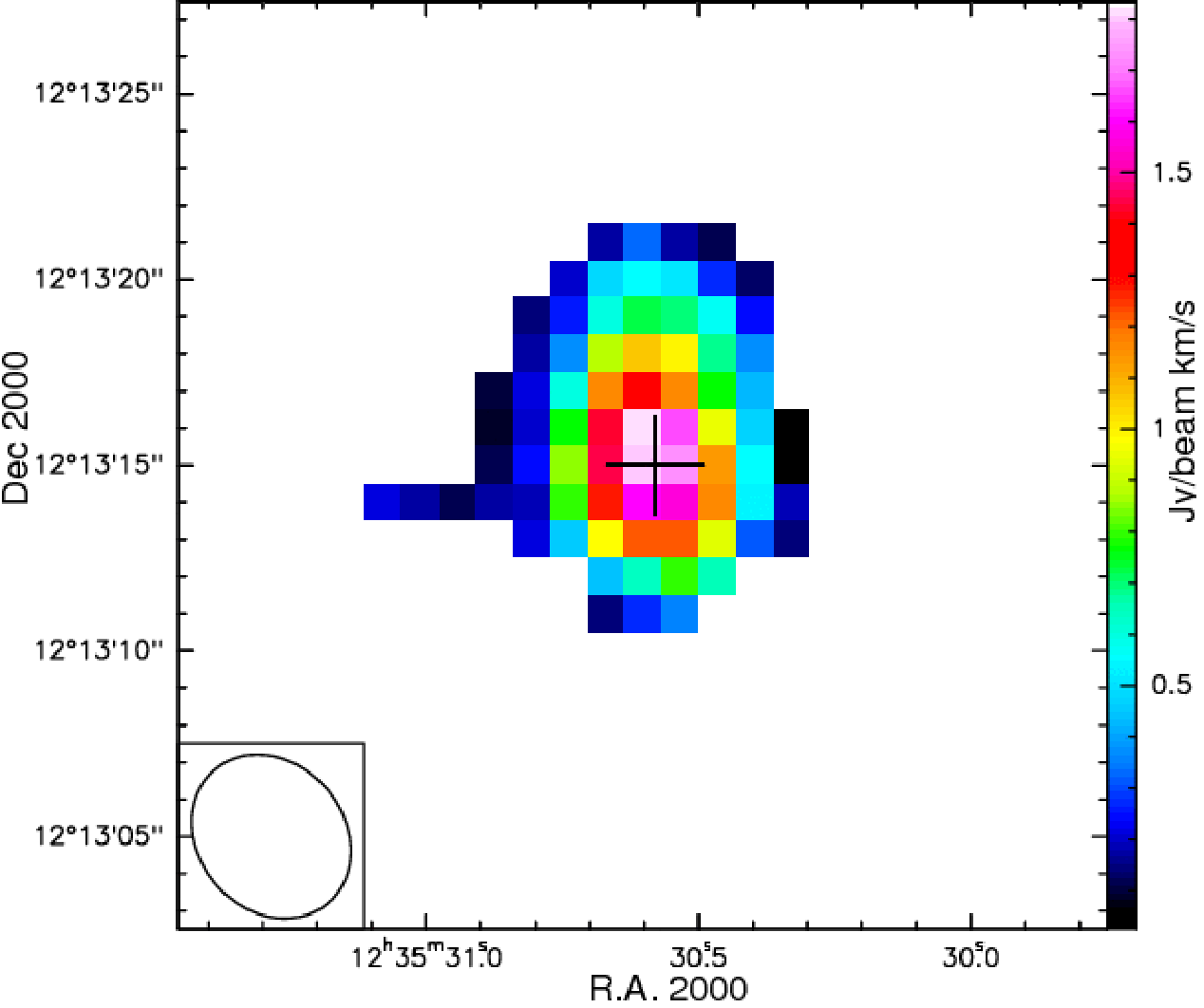}}
\hspace{5mm}
\rotatebox{270}{\includegraphics[width=7cm]
  {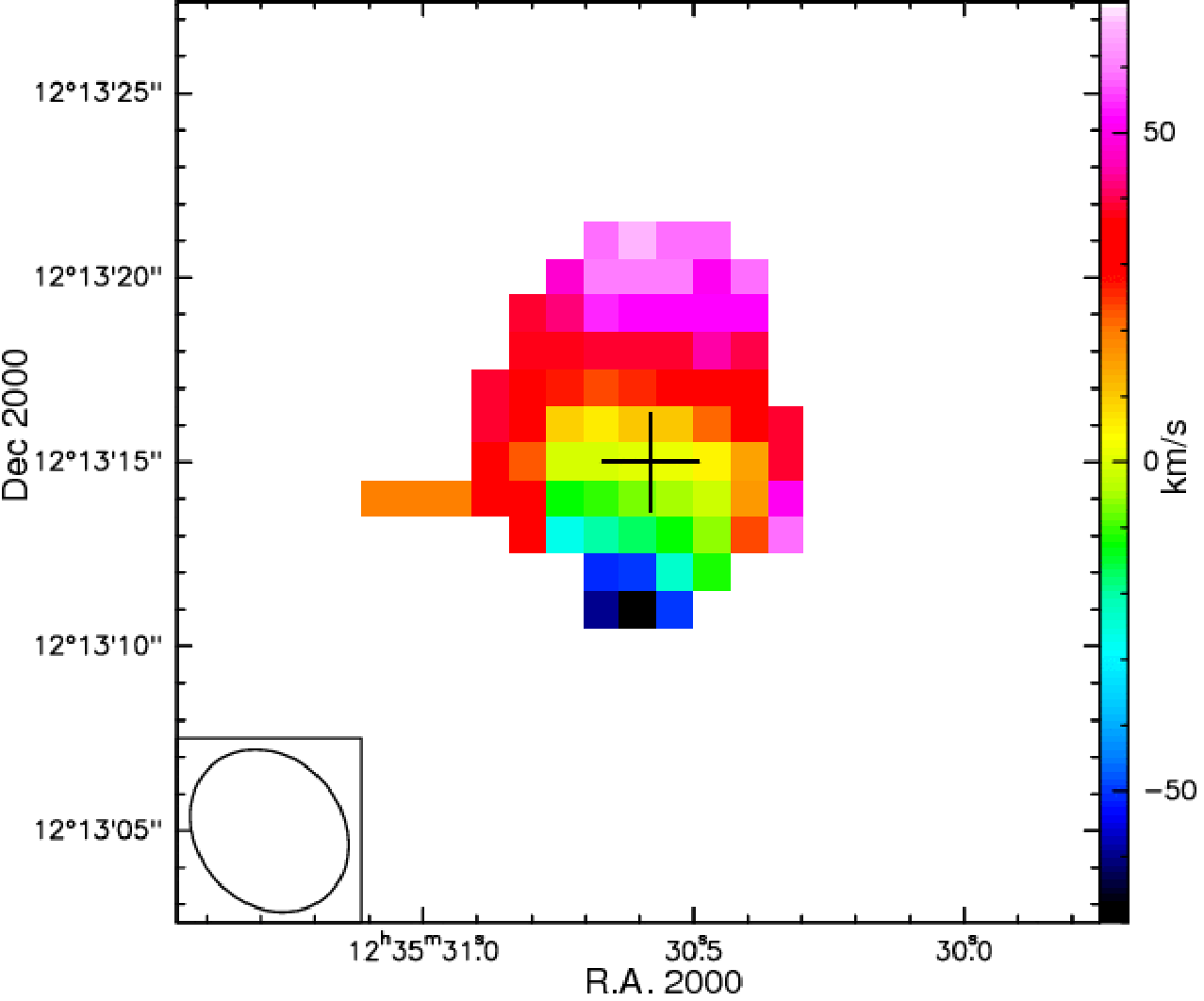}}\\
\caption{ {\em Left:} CO(1-0) integrated intensity map of NGC~4550. {\em
    Right:} CO(1-0) mean velocity map relative to the systemic
  velocity of 435 km s$^{-1}$ (see Table 1). The synthesized beam is shown in the
  bottom-left corner of each map. The cross represents the centre of
    the galaxy as given by 2MASS. \label{fig:meanvelo}}
\end{center}
\end{figure*}

NGC~4550 was observed in the  $^{12}$CO(1-0) line with the
IRAM Plateau de Bure Interferometer (PdBI) on August 10, 2007. The
observations were taken in the D configuration with 5 antennas and used the new
generation dual-polarization receivers at 3~mm. The spectral
correlators were centred at 115.12~GHz, the transition frequency of
CO(1-0) roughly corrected for the heliocentric velocity of NGC~4550.  The
correlator configuration used four 320~MHz-wide (834 km s$^{-1}$)
units with a frequency resolution of 2.5~MHz (6.6 km s$^{-1}$),
covering a total usable bandwidth of 950~MHz (2475 km s$^{-1}$). The
correlator was regularly calibrated by a noise source inserted in the
IF system.   

We obtained visibilities with series of thirty 45 s integrations on
source, followed by three $45$ s phase and amplitude calibrations on
each of the two phase calibrators (see Table 2). To flux calibrate, we
set the flux density of 3C273 to the expected value (15 Jy), then
checked that the fluxes found for the other flux calibrators were
reasonable. The uncertainty in our flux calibration is $\approx20$\%. 

The data were reduced with the Grenoble Image and Line Data Analysis
System (\textsc{gildas}) software packages \textsc{clic} and \textsc{mapping} \citep{GL}.
We calibrated the data using the standard pipeline in \textsc{clic}. 
  After calibration, we used \textsc{mapping} to
create data cubes centred at 115.104~GHz (corrected more precisely for 
a systemic velocity of 435 km s$^{-1}$) with velocity planes separated by 20 km
s$^{-1}$. Natural weighting was used. The primary beam size is $44\arcsec$ for
CO(1-0) and we chose the spatial dimensions of the datacube to be
about twice the diameter of the primary beam,
$90\arcsec\times90\arcsec$.  The synthesized beam is $4\farcs7 \times
4\farcs0$, so we choose spatial pixels of
$1\arcsec\times1\arcsec$. Continuum subtraction was unnecessary (see
below). The dirty beam has small
side lobes that necessitated cleaning the datacube. The cleaning was
done using the H\"ogbom method \citep{hogbom74}; we stopped cleaning
in each velocity plane after the brightest 
residual pixel had a value lower than the rms noise of the 
uncleaned datacube. 

To constrain the continuum emission, we selected frequencies at least 40
km s$^{-1}$ away from the lowest and highest velocity channels with any
line emission in the cleaned datacube (-80 and 80 km s$^{-1}$,
respectively). The very edges of the bandwidth were also
avoided, resulting in a 683 MHz wide continuum window. We mapped
this data with the same spatial parameters as the line data. No
emission could be seen in the map with an rms noise of 0.54 mJy beam$^{-1}$,
giving a 3$\sigma$ upper limit of 1.62 mJy assuming a point source. 

\subsection[]{Results}

The channel maps (Fig.~\ref{fig:chmap10}) show CO(1-0) emission
around the galaxy centre. The emission is just over
the 3$\sigma$ level, with stronger emission in the
20, 40 and 60 km s$^{-1}$ channels. The channel velocity values are
all given with respect to systemic velocity. In the channel maps, a mild
velocity gradient is observed, with more of the emission at positive
relative velocities to the north of the galaxy centre and more of the
emission at negative relative velocities to the south. The 60 km
s$^{-1}$ channel is an exception with a peak slightly south
of the galaxy centre, although the emission clearly also extends to the north. 
To pick up more extended emission, we created an integrated intensity
map and a mean velocity field map (Fig.~\ref{fig:meanvelo}) using the
smoothed-mask 
method. To make these maps, we first created a smoothed cube by
smoothing with a 2D circular gaussian spatially (FWHM of 4\arcsec or
about one synthesized beam width). Then we hanning
smoothed by 3 channels in velocity. The moments were computed
only using pixels in the original cube that corresponded to pixels
above 3$\sigma$ in the smoothed cube.

\begin{figure}
\begin{center}
\rotatebox{270}{\includegraphics[width=7cm]
{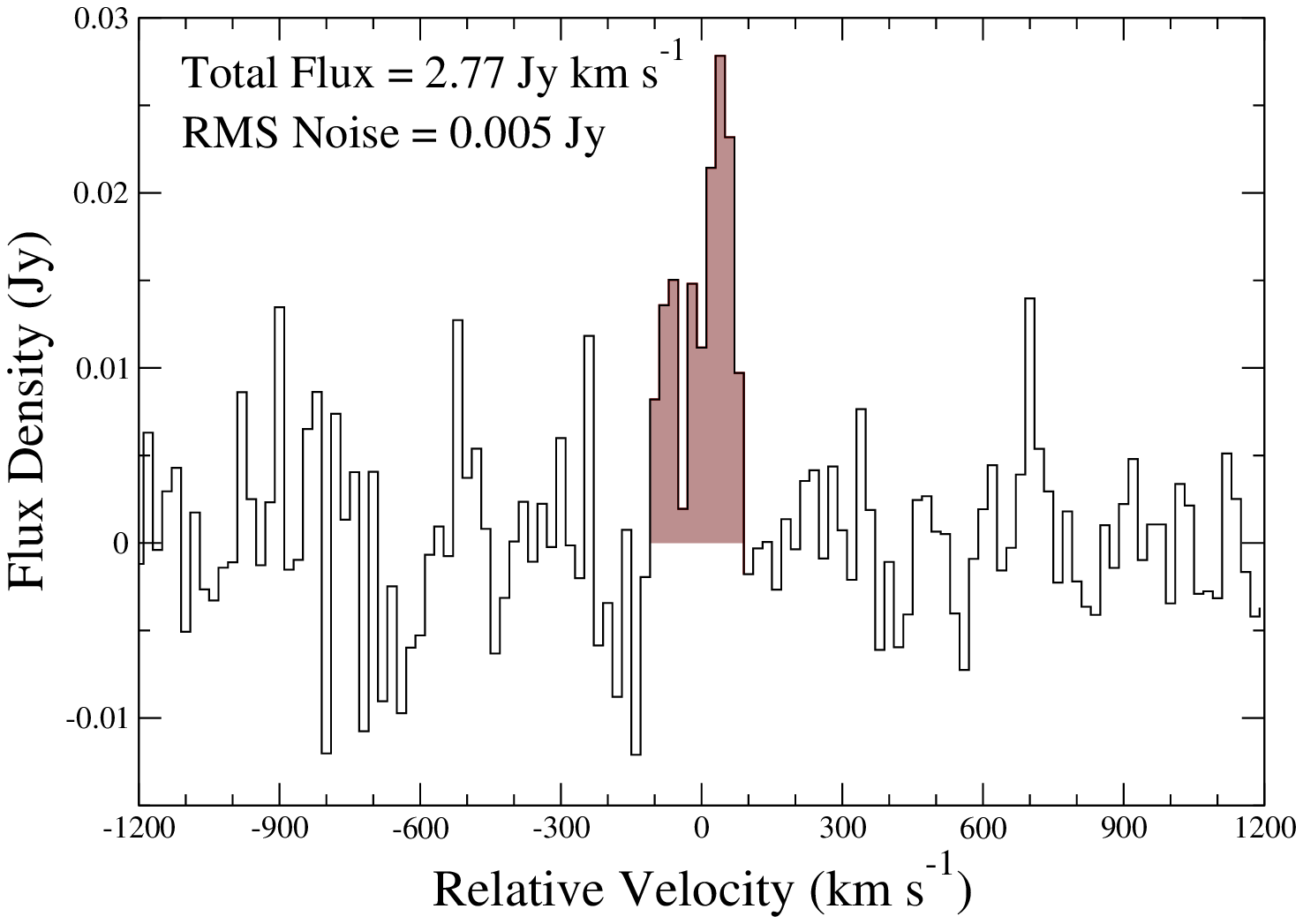}}
\caption{Spectrum of the spatial region with observed CO(1-0)
  emission in NGC~4550.\label{fig:spec} }
\end{center}
\end{figure}

The integrated intensity map shows that the emission is limited to the
central 750 pc (10\arcsec) in diameter and is slightly stronger in the north. 
The mean velocity map shows a north-south
gradient, as expected from the shift observed in the channel
maps. Assuming the velocity gradient indicates rotation around
the centre of the galaxy, the molecular gas rotates like the observed
ionized gas and the thicker disc. Spatially integrating the
entire datacube over the region with 
observed emission results in the spectrum shown in
Fig.~\ref{fig:spec}. This spectrum shows that the emission is present
in all channels between -80 and 80 km s$^{-1}$, but is strictly
limited to these channels.

\begin{figure}
\begin{center}
\rotatebox{0}{\includegraphics[width=7cm]{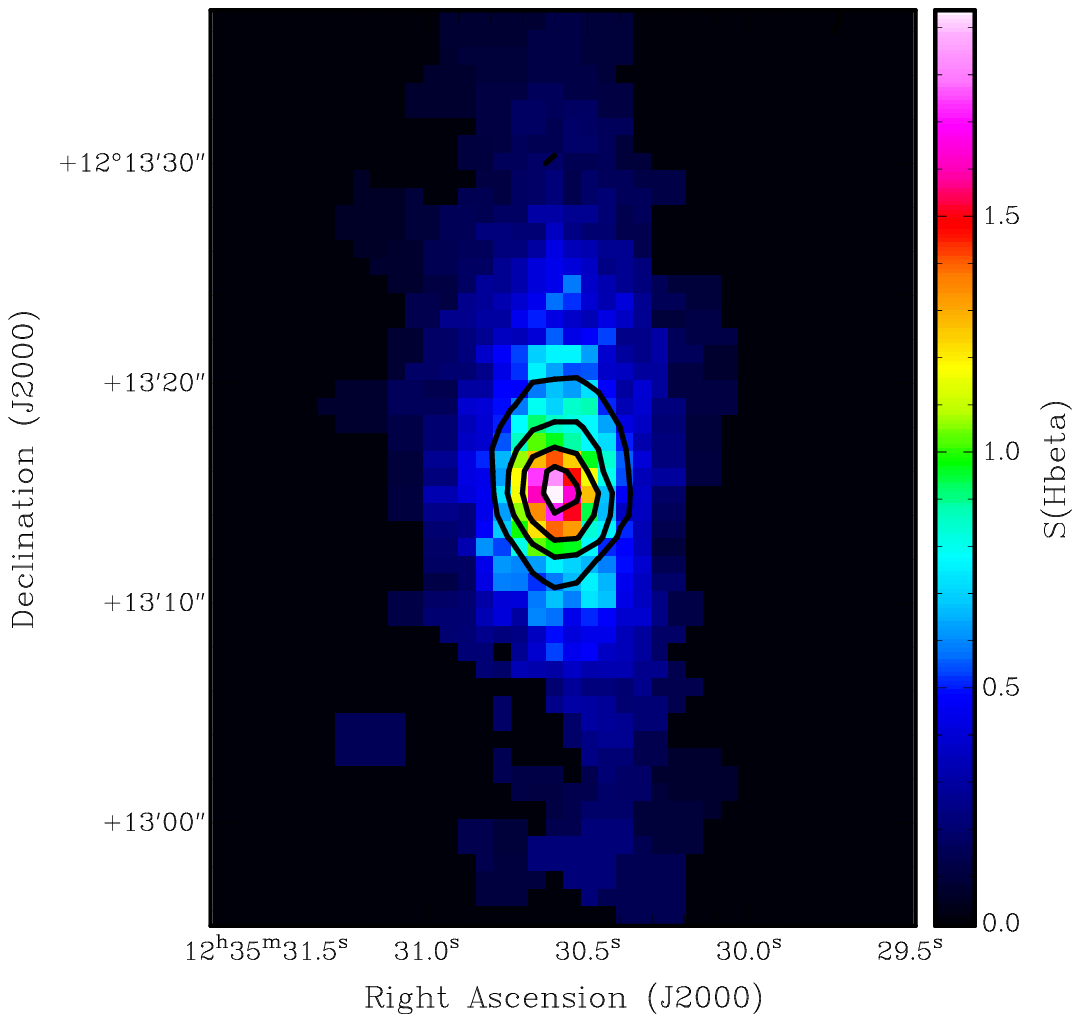}}
\caption{CO(1-0) contours over the \hbeta emission map of NGC~4550 from SAURON
  \citep{sarzi06}. The \hbeta 
  emission is clearly more extended than the CO. CO(1-0)
  contours are at 0.25, 0.75, 1.25 and 1.75 Jy beam$^{-1}$ km s$^{-1}$.\label{fig:ionized} }
\end{center}
\end{figure}

While the lopsidedness of the CO emission was discovered
by \citet{wiklind01}, the degree of asymmetry was exaggerated due to
their use of  an incorrect systemic velocity. The
systemic velocity listed in NASA Extragalactic Database (NED) for NGC~4550 and that used by
\citet{wiklind01} is $381 \pm 9$ km s$^{-1}$. This value is based on
the H{\small{I}} measurement of \citet{peterson79}. More recent and more
sensitive H{\small{I}} observations have not detected H{\small{I}} (DuPrie
\& Schneider, 1996; Morganti et al., in preparation), making the
\citet{peterson79} value very questionable. In addition, recent
absorption line studies seem to be converging to a value of around 435~km~s$^{-1}$ (\citet{rubin97} find 434~km~s$^{-1}$, \citet{uzc00} list
$458 \pm 41$~km~s$^{-1}$, \citet{wegner03} find $437 \pm
15$~km~s$^{-1}$ and the velocity in the central pixel of the SAURON
velocity map is $435$~km~s$^{-1}$). We have adopted this value for our
work. Then, the 
flux from positive relative velocities is 1.64 
Jy km s$^{-1}$, around twice as much as the flux from negative
relative velocities of 0.91 Jy~km~s$^{-1}$. The ionised gas
does not show the same bias as the molecular gas, with fairly
symmetric emission in both space and velocity. In addition, the
extent of the ionised gas is much greater than that observed in the
molecular gas (see Fig.~\ref{fig:ionized}). As shown in
\citet{wiklind01}, however, the dust favours the north, in agreement with the molecular gas bias.

Integrating the spectrum over the velocity range with observed
emission, i.e. from -80 to 80 km s$^{-1}$, we obtain a total CO
flux of 2.77 Jy km s$^{-1}$. 
The total flux is used to compute the total molecular
 hydrogen mass using the formula $M(\mathrm{H}_2) = (1.22 \times 10^4
 \mathrm{M}_{\sun})D^2 \times S_{\mathrm{CO}}$, where D is the distance
 measured in Mpc and $S_{\mathrm{CO}}$ is the total CO(1-0) flux. This
 formula comes from using the standard CO to H$_{2}$ conversion
 ratio $N(\mathrm{H}_2)/I(\mathrm{CO}) = 3 \times 10^{20}$ cm$^{-2}$,
 where $N(\mathrm{H}_{2})$ is  the column density of H$_{2}$ and
 $I(\mathrm{CO})$ is the CO(1-0) intensity in  K km s$^{-1}$. We note
 that the actual conversion ratio for NGC~4550 is unknown and can be
 expected to vary by a factor of at least two
 \citep[e.g.][]{leroy07}. However, using the specified conversion 
 ratio, the
 total H$_{2}$ mass  detected by our interferometric observations is
 $8.1\times10^{6}$ M$_{\sun}$. 

Using the same value of $N(\mathrm{H}_2)/I(\mathrm{CO})$ and the same
distance to NGC~4550, the single-dish measurement of \citet{wiklind01}
gives a molecular mass of $1.4 \times 10^{7}$ $\rm M_{\sun}$, slightly
less than twice the value we obtain. However, they integrate over a
much larger velocity range (225 to 535 km s$^{-1}$), likely including
much noise as positive emission (our velocity range is only 355 to 515
km s$^{-1}$). Over a similar velocity range, our estimates for the
molecular mass in NGC~4550 would be in suitable agreement, considering
calibration uncertainties ($\approx20$\%). Reflecting the calibration uncertainty and
the much larger uncertainty in the conversion ratio, we report the
molecular mass in NGC~4550 as $1\times 10^{7}$ $\rm M_{\sun}$,
which is the value we will use for the rest of the paper.  

\begin{figure*}
\centering
\includegraphics[width=16.4cm,clip]{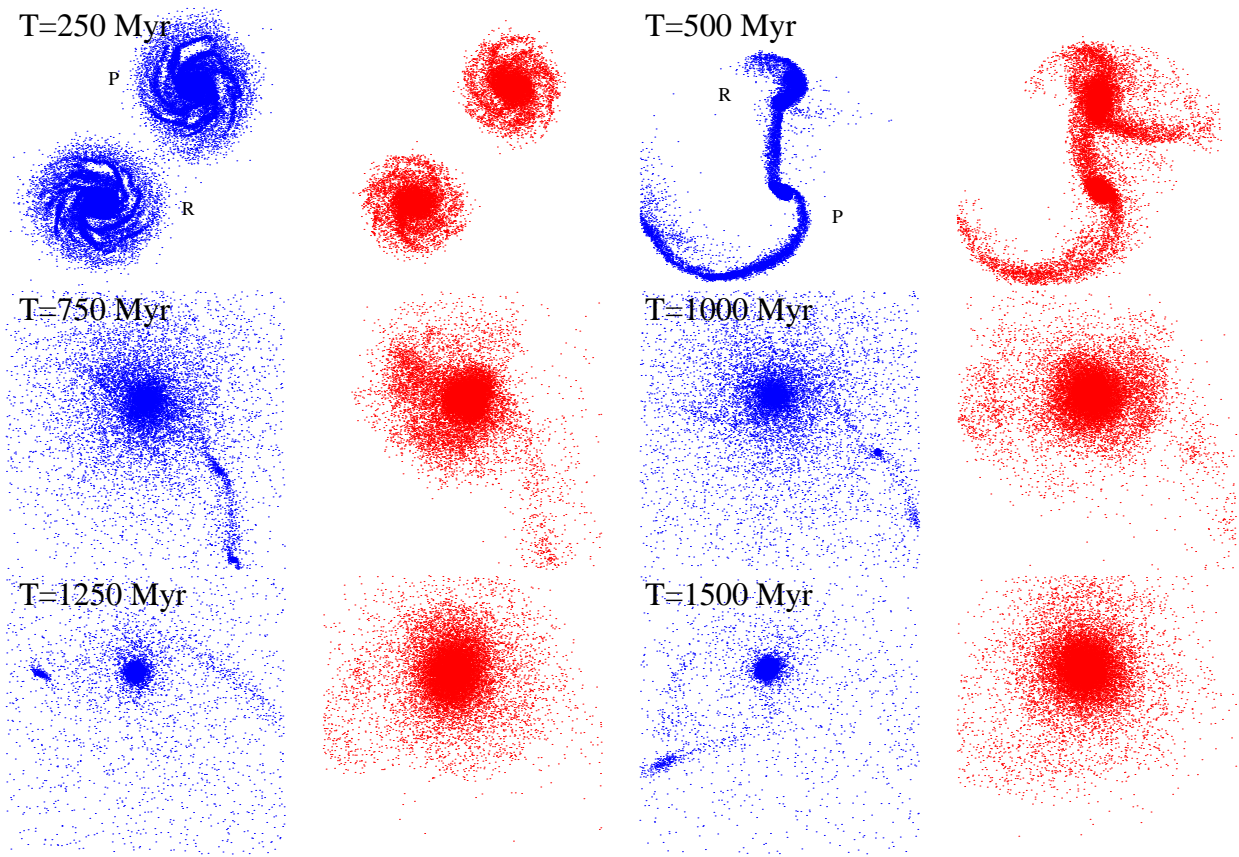}
\caption{Particle plots (x-y projection) of six snapshots of the
  simulation: the face-on views of the gas (left, blue) and stars
  (right, red) are plotted for each epoch, T=250 to
  1500 Myr in steps of 250 Myr.  The relative orbital motion is
  counter-clockwise in this picture, where the prograde galaxy is labeled "P" and the
retrograde one "R". The merger is complete at the 3rd snapshot
  (T= 750 Myr), with some gaseous tidal dwarfs remaining
  thereafter. All boxes are 80~kpc wide. }  
\label{pplot}
\end{figure*}

\begin{figure*}
\centering
\includegraphics[angle=-90,width=16.4cm,clip]{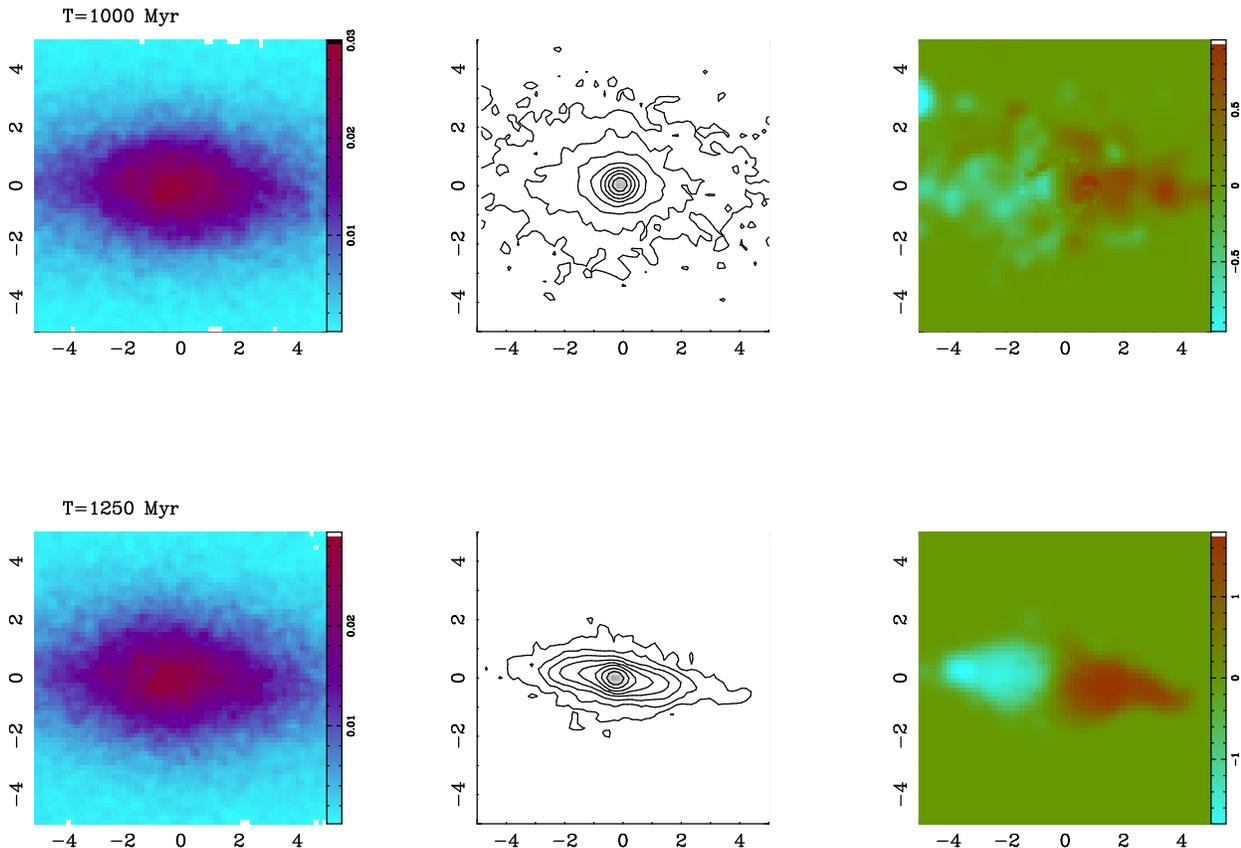}
\caption{The edge-on (x-z) stellar density (left), gas density
  contours (middle) and gas velocity field (right) are plotted
  for two times after the merger, T=1000 and 1250~Myr. It is easy
  to see that the settling of the gas occurs during this epoch: the
  gas density contours flatten and the gas velocity field becomes
  regular. The boxes are all 10~kpc
  wide, the wedges at the right of the velocity maps indicate the amplitude
  of the projected  velocities in unit of 100~km~s$^{-1}$.  All later
  snapshots are similar to the T=1250 Myr one, with a regular
  rotation velocity map for the gas.} 
\label{isovels}
\end{figure*}

\begin{figure*}
\centering
\includegraphics[angle=-90,width=16.4cm,clip]{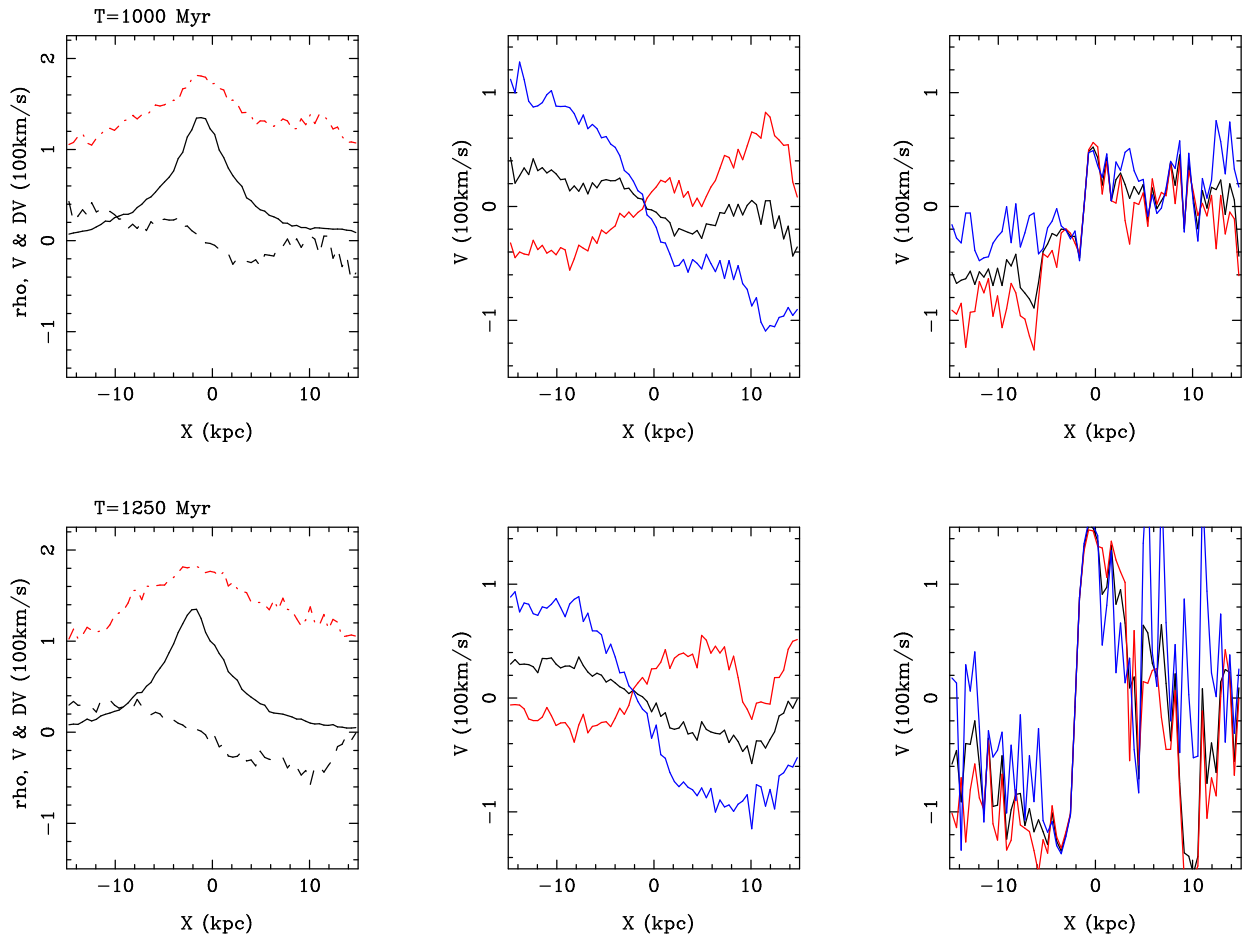}
\caption{Velocity evolution during the simulation. For the same epochs as Figure
6, the left panel shows the stellar density (solid black line), mean
line-of-sight velocity (black dashed line) and velocity dispersion
(dot-dashed red line) profiles. The central panel shows the mean
stellar velocity profile of the resulting system again (solid black
line), as well as those of the prograde (solid red line) and
retrograde (solid blue line) galaxies individually.  The right panel
is as the central panel but for the gas velocities. All the profiles
were extracted from a 4.5~kpc thick horizonthal slice taken from
the edge-on projection shown in Figure 6. All boxes are 30~kpc wide.} 
\label{rotcurv}
\end{figure*}

\section[]{Numerical Simulation}

While both the gas accretion and coplanar major merger scenarios seem
to be able to produce at least the general properties of the two
stellar discs in NGC~4550, we would also like to know how the gas in
this galaxy came to rotate like the thicker disc. Gas rotating
like the thick disc is actually in disagreement with the
predictions of the gas accretion scenario, in which the thin disc is
formed from the acquired cold gas \citep{kenney00}. Any remaining gas would then be
expected to rotate like the thin disc. To see what the merger scenario
predicts for the gas, we have performed a numerical simulation,
described below. 

\subsection[]{Method and initial conditions}

 Our merger simulation for NGC~4550 has been run with the same
 TREE-SPH technique 
 as described in Di Matteo et al. (2007).  The two merging galaxies
 are initially of the same type (Sb) and mass (each has a total mass
 $2.4 \times 10^{11}$ M$_\odot$ including dark matter).  
The bulges and halos are modelled as Plummer spheres, of masses 1.15
 and $17.25 \times 10^{10}$ M$_\odot$, and characteristic radii 1 and
 12~kpc, respectively.  The stellar discs have masses of $4.6
 \times 10^{10}$ M$_\odot$ and radial scale lengths of 5 kpc. The
 gas to stellar  disc mass ratio is 0.2. 

The initial separation between the two galaxies is 100 kpc. We choose
an orbit for the galaxies such that they would have a parabolic
encounter if they were both point masses. If the galaxies followed the
parabolic orbit,  the first pericentre would be 8~kpc and
their relative velocity at this point would be 707 km s$^{-1}$. Of
course, the model galaxies do not follow this keplerian
orbit. Instead, given the strong dynamical friction,
merging occurs rapidly.  The orbital angular momentum is initially
oriented in the positive z-axis as is the spin of one of the two
galaxies (hereafter called the prograde galaxy). The other galaxy has
negative spin (and will be called the retrograde galaxy). 

The total number of particles is 240 000, with 120 000 in each galaxy,
divided between 40 000 in the gas, 40 000 in the stars and 40 000 in the
dark matter halo. We use the Schmidt law and hybrid particles to take
star formation into account as in Di Matteo et al. (2007). 
 In the following, when we refer to the  ``gas component'', this means
 the sum of the gas particles and the very young stars formed since the merger.

\subsection{Analysis and Results}

Figure \ref{pplot} shows the gas and stellar particles for six
snapshots from 250 to 1500~Myr. The simulation was run until T=4000
Myr (with a timestep of 0.5 Myr), but the evolution after 1500~Myr is
only secular relaxation of the merger remnant. 

With the spin of one galaxy aligned with the orbital motion and the
spin of the other anti-aligned, the interaction is strongly asymmetric
despite the 
equal mass of the galaxies. The prograde galaxy develops long tidal
tails, while the retrograde is less heavily influenced by the interaction. As the
merger evolves, the central regions become dominated by the stars from
the retrograde galaxy as it remains more compact. However, the gas behaves
differently due to its collisional nature. While two counter-rotating
systems of stars can coexist, one of the gas systems must  
dominate. Initially, the gas from the prograde galaxy expands in tidal
tails and even tidal dwarfs. However, this gas then falls back and
settles in the disc of the merger remnant due to its dissipative
nature. Since the orbital angular momentum is positive, the prograde
gas increases its angular momentum and the retrograde gas loses some
of its angular momentum in the encounter. This results in the gas
rotating with positive angular momentum after the gas from the two
discs has interacted and fully settled down.  

Figure \ref{isovels} displays both the edge-on gas density and
velocity map for two time steps: it explicitly reveals how the gas settles 
into a thin disc while re-ordering its kinematics to prograde  
rotation. More quantitatively, Figure  \ref{rotcurv} shows the global
rotation curves along an edge-on projection, splitting the
rotational profiles according to the various components, the gas and
stars in the two galaxies. The rotation curves make it very clear that
the stars in the retrograde galaxy retain their direction of rotation,
dominating the total rotation curve in the centre of the galaxy. The
stars from the prograde galaxy show a lower amplitude of rotation, as
they have been strongly heated in the encounter. 
The merger has thus created a disc galaxy with one dynamically hotter disc of
stars, one dynamically cooler counter-rotating disc of stars and a gaseous
component rotating like the thick disc. As the alignment of the
orbital angular momentum with the prograde galaxy causes both the
heating of this galaxy and the settling of the gas into prograde
rotation, we predict that in such systems the gas will always be in
corotation with the  
most perturbed stellar system, i.e. with the hotter disc. 
NGC~4550 clearly exhibits this phenomenon with both the molecular and
ionised gas rotating like the thicker stellar disc. 

However, in the present simulation with perfectly coplanar galaxies, we note that there is not much heating in the perpendicular direction. The thickness of the stars with positive angular momentum (those that would observationally be considered in the prograde disc) is comparable to the thickness of the stars with negative angular momentum. While the lack of a bona fide thicker disc limits the direct comparison to NGC~4550, we expect more resonant heating in the perpendicular direction if the interaction were not perfectly coplanar. Earlier simulations show that the angle between the discs' orientation and their orbit must be significant before the remnant takes on an elliptical morphology \citep{bournaud05}, leaving room for disc thickening before disc destruction. We will explore the possible parameters of low-inclination mergers in a future work. 

\begin{figure*}
\centering
\includegraphics[angle=-90,width=15.5cm,clip]{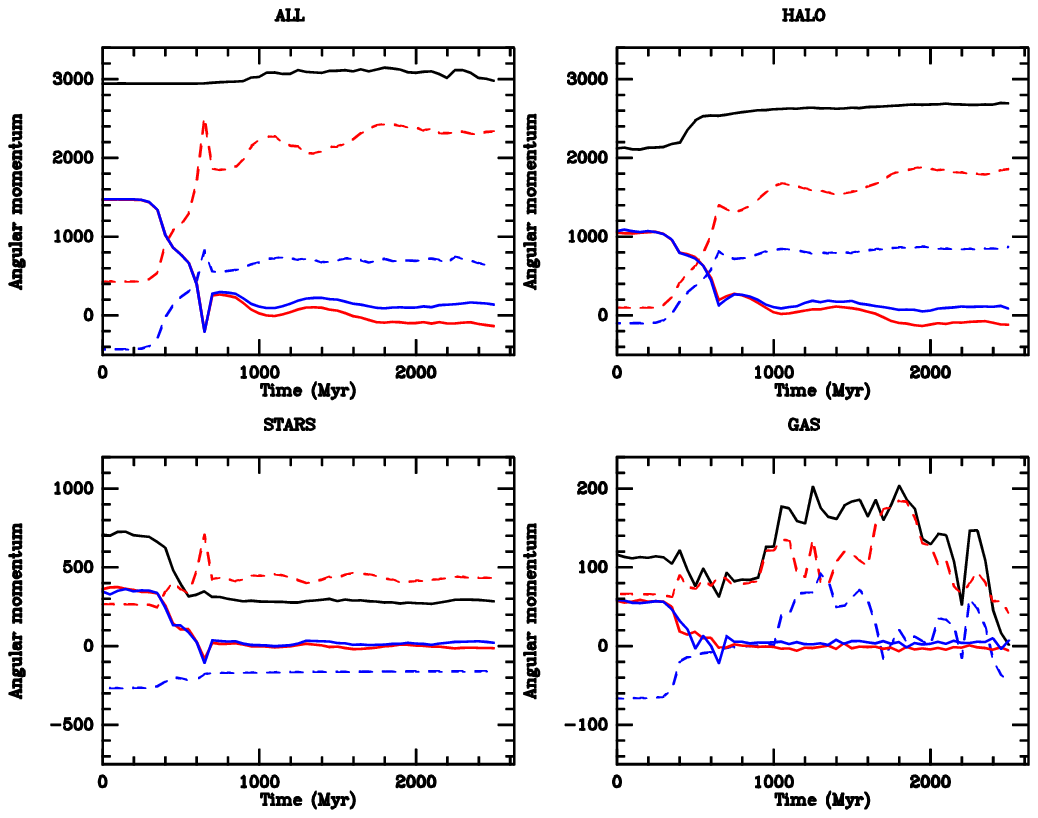}
\caption{Angular momentum evolution during the simulation.
The top left panel shows the sum of all components, while the other three
are split into the dark matter halo, the stars and
the gas, as indicated above each panel. The wiggles in the total
angular momentum line in the upper left box reflect noise in the
simulation of order 2-3\%. The colour scheme is as for
middle and right panels of Figure 7, i.e. red for the prograde galaxy
and blue for the retrograde galaxy. The solid lines indicate the
orbital angular momenta, while the dashed lines indicate the internal
spin momenta. The angular momentum is in units of $2.3 \times 10^{11}$ M$_\odot$
kpc km s$^{-1}$; note that the scale is different in each plot. }  
\label{angmom}
\end{figure*}

As angular momentum exchange strongly drives the evolution of the
merging galaxies, it is interesting to follow the angular momentum
exchange between all components present, including the dark matter.  
Figure \ref{angmom} shows clearly that the largest angular momentum
exchanges occur during the violent merging epoch between 
400 and 800 Myr. Afterwards, all momenta evolve more slowly.
  The main feature is the loss of all the initially positive 
orbital angular momentum in each component, which is then transferred
almost exclusively to the dark matter, except for a tiny fraction
transferred to 
the internal spin of the gas and stars.
 This exchange is seen in the total global momentum (black full curve)
in each panel of Figure \ref{angmom}. It is particularly intriguing to
note that both dark haloes acquire a large prograde rotation, an
effect even more marked for the halo of the prograde galaxy.

Our simulation agrees with the results of Puerari \& Pfenniger
(2001) who first showed that a coplanar major merger
between two disc galaxies could limit the heating effect and create a
NGC~4550 look-alike. In their simulation, they consider either a
parabolic or a circular orbit, assuming that the relative energy of
the two galaxies coming from infinity has already been absorbed by
some outer matter. Both orbits avoid excessive heating, but only
the parabolic orbit produces the spectacular counter-rotation
observed in NGC~4550. However, their simulations are limited in how
they investigate the gas- they include gas in either the prograde
galaxy or the retrograde galaxy, but do not consider the case where
gas is present in both galaxies as we have done. Of course, the final
distribution of the gas 
within the merger remnant will depend on its initial distribution in
the two progenitor galaxies.

\citet{dimatteo08} have also studied dynamical mechanisms
to produce counter-rotating systems, through the merging of
a non-rotating  early-type galaxy with a gas-rich spiral. In these
types of mergers, the orbital momentum is transferred to the rotation
of the early-type galaxy and to the outer parts of the spiral, while
the centre of the spiral (gas and stars) 
keeps its retrograde momentum. The counter rotation is
then spatially marked (centre versus the outer parts), which is a
different pattern than that observed in NGC~4550, where gas and stars are
counter-rotating at the same location. The nature, amount and extent
of the observed counter-rotation can thus help to determine
the merging history of the system.

\section[]{Star Formation}

The CO detected in NGC~4550 indicates the potential for star
formation, as stars originate in cold molecular gas. Using the Toomre
density criterion, Q, with a conservatively cold velocity dispersion of
6~km~s$^{-1}$ and an epicyclic frequency measured from the rotation of
the ionised gas \citep{sarzi06}, we calculate a threshold density
(Q=1) of 285 \msun 
pc$^{-2}$. Determining a gas surface  
density for NGC~4550 from the CO data is difficult given the lack of
resolution, but assuming a uniform-density, symmetric
5\arcsec-radius disc, we obtain a surface density of only 18.3 \msun pc$^{-2}$ for
the molecular gas, far below the threshold. Even decreasing the radius
to 2\arcsec, about the minimum believable extent of the molecular gas,
gives a surface density of only 115 \msun pc$^{-2}$, still below
the star formation threshold density. However, if the molecular gas
is significantly more clumpy than a uniform density disc, than there
may be isolated regions where the critical density is reached. We
therefore do not immediately rule out the possibility of star
formation in NGC~4550, instead looking for signs in other star
formation tracers.


Far-infrared (FIR) emission is often used as a tracer of star
formation, as young OB stars efficiently heat their surrounding
dust. In late-type galaxies where the ultraviolet (UV) and visible
radiation is dominated by young stars, the FIR emission is a very good
tracer of  star formation. Its use in earlier-type galaxies, however,
is more disputable, as the radiation field of old hot stars should also
contribute to the dust heating. Yet the relations between other
tracers of star formation such as H$\alpha$ \citep[e.g.][]{kewley02} and radio
continuum \citep[e.g.][]{gavazzi86}  hold puzzlingly constant with
morphology (at least within spirals), indicating that it is dust proximate to star forming
regions which  dominates the FIR emission. Converting the IRAS 60 and
100 $\mu$m fluxes into a star formation rate (SFR) following the prescription of
\citet{kewley02}, we derive a FIR-based SFR of $\approx0.017$ \msun
yr$^{-1}$ for NGC~4550.

\begin{figure}
\begin{center}
\rotatebox{0}{\includegraphics[width=6cm]
  {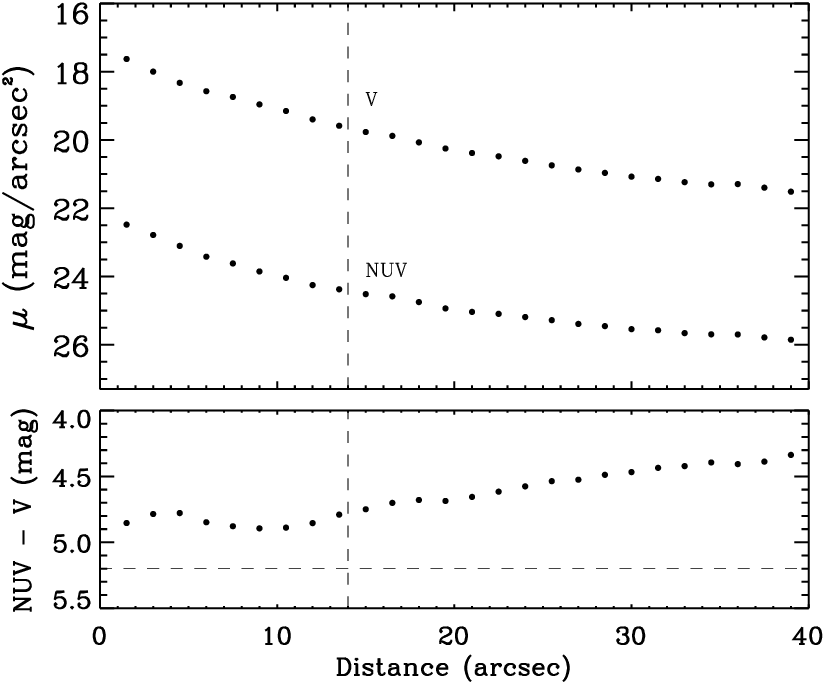}}
\caption{NUV and V surface brightness profiles and NUV-V colour along
  the major axis in NGC~4550. The vertical dashed line indicates the
  effective radius while the horizontal dashed line indicates the blue
  limit to the colours the uv-upturn phenomenon can produce,
  NUV-V=5.2. \label{fig:nuv-v}}
\end{center}
\end{figure}

As many early-type galaxies without star formation also emit in the
FIR (an active
galactic nucleus or infrared cirrus can also contribute), we do not
want to rely on the FIR emission alone to ascertain ongoing star
formation in NGC~4550. Instead, we look to the UV, where young stars
are particularly bright. The first diagnostic we consider is the NUV-V
colour. Using optical data from the MDM 1.3m telescope 
(Falc\'on-Barroso et al. in preparation) and {\it   GALEX} UV data 
(Jeong et al. in preparation), we have produced a plot of the NUV-V colour
along the major axis of NGC~4550
(Fig.~\ref{fig:nuv-v}). NGC~4550 has a central NUV-V colour of around 
5, becoming bluer with increasing radius. While old stars can also
emit in the UV, producing the UV-upturn phenomenon seen in old
galaxies \citep[e.g.][see O'Connell 1999 for a review]{burstein88}, the bluest NUV-V colour this effect
produces is NUV-V$=$5.2 \citep{yi05}. Thus, the NUV-V colour in NGC~4550 is
more likely from young stars. Furthermore, the
decreasing gradient (bluer colour with increasing radius) is not seen
in UV-upturn galaxies, reinforcing the idea that the NUV emission
comes from young stars. 

\begin{figure*}
\begin{center}
\rotatebox{0}{\includegraphics[width=6cm]
  {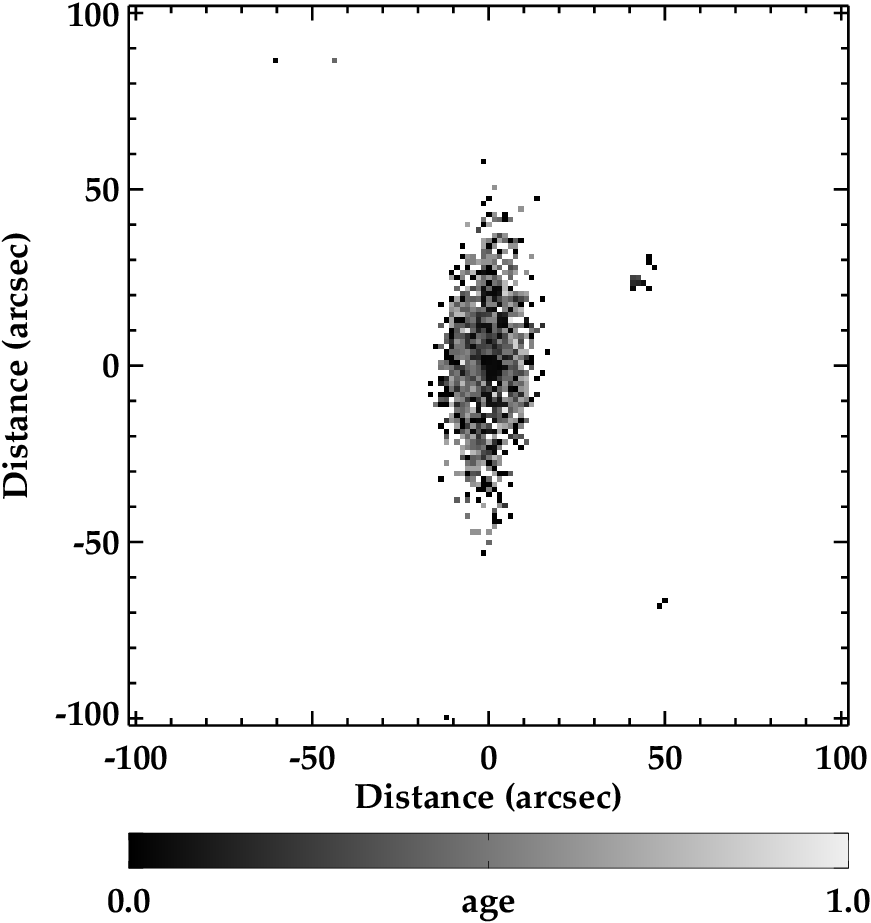}}
\rotatebox{0}{\includegraphics[width=6cm]
  {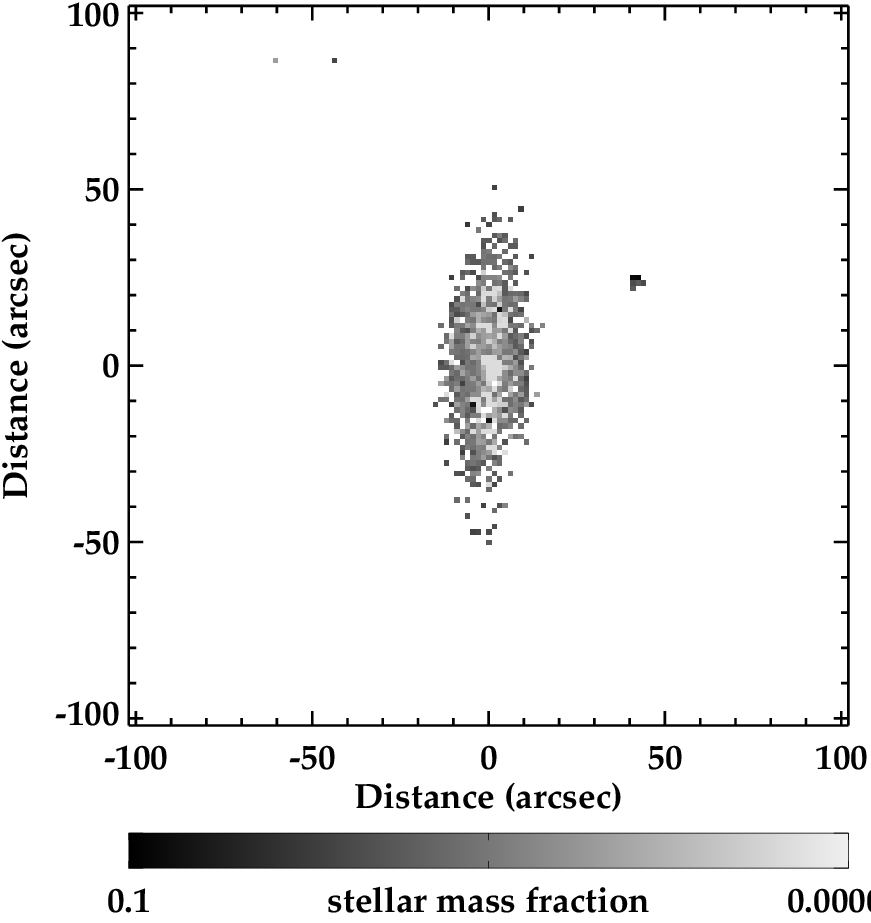}}\\
\caption{ 
{\em Left:} Two-component fit map of the age
    of the young component (in Gyr).  {\em Right:}
    Two-component fit map of the mass fraction of the young
    component. Note that the mass fraction is relevant to
    each individual pixel, thus a certain mass fraction in the central
    pixels represents a larger mass of young stars  than the same mass
    fraction in outer pixels. \label{fig:uv} }
\end{center}
\end{figure*}

 
\begin{figure*}
\begin{center}
\rotatebox{0}{\includegraphics[width=14cm]{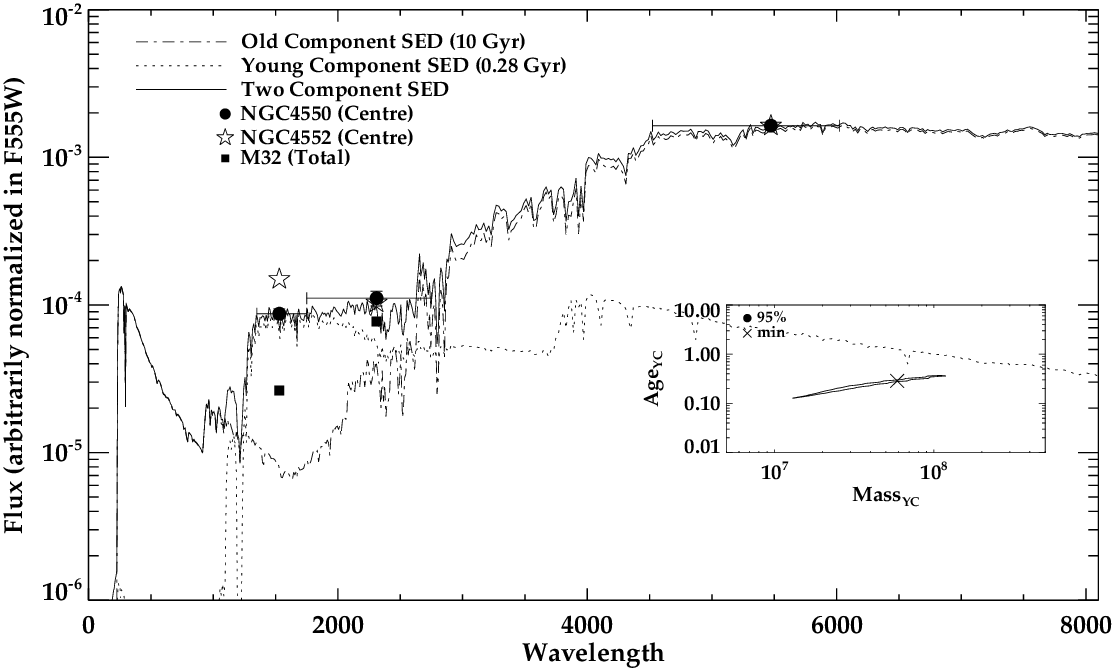}}
\caption{Integrated V, NUV and FUV fluxes for the central region of
  NGC~4550. The central region is defined by an ellipse
  of 5\arcsec$\times$3\arcsec, PA=0$^{\circ}$. The spectral energy
  distributions (SEDs) of the best-fit young and old components are shown,
  along with the SED of the sum of these components. We have also
  shown the integrated 
  colours of NGC~4552 (strong UV-upturn galaxy) and M32
  (intermediate-aged compact elliptical) normalized in the V-band for
  comparison. {\em Inset:} 
$\chi^{2}$ contours of the two-component stellar
  population fit to the central integrated colours of NGC 4550,
  showing the age-mass fraction degeneracy. The best
  fit is marked with an 'x'.\label{fig:chicont} }
\end{center}
\end{figure*} 

To investigate this possibility, we have used the method
described in \citet{jeong07} to fit a two-stage star formation history to the
UV-optical colours of NGC~4550. An old population is fixed at an
age of 10~Gyr with a composite metallicity, while a younger component at 
solar metallicity is allowed to vary in age ($0.001$ Gyr $<$ {\it
  t}$_{\mathrm{YC}}$ $< 10$~Gyr) and mass fraction ($10^{-4}<$
{\it f}$_{\mathrm{YC}} < 1$).  
Figure~\ref{fig:uv} shows the results of these
fits, pixel by pixel. The central region appears to host a young stellar population
(fitted age around 100~Myr), although the mass fraction is very low at only
$~$0.01\%. To increase the signal-to-noise and thus the reliability of
our model fits, we computed integrated colours for the central
region where the CO emission is detected, using an ellipse with a major
axis of $5\arcsec$, a minor axis of $3\arcsec$ and a position angle of
0$^{\circ}$. The colours from this aperture give a best-fit age of
280~Myr with $5.89 \times 
10^{7}$~M$_{\sun}$ of young stars. The fit to the spectral energy distribution (SED)
of this central region of NGC~4550 is shown in
Figure~\ref{fig:chicont}. As seen in the figure, NGC~4550 is slightly 
bluer in NUV-V than the strong UV-upturn galaxy NGC~4552 and much bluer than
the intermediate age compact elliptical M32, indicating that young
stars must be responsible for the blue NUV-V
colour of NGC~4550. Figure~\ref{fig:chicont} also displays an  
inset of the chi-square contours for the fit. As can be seen, an
age-mass degeneracy is present, with larger mass of
slightly older stars (but only up to 500~Myr) or a smaller mass of even younger stars also able to fit
the colours.   


Absorption linestrengths also give an indication of a young or
intermediate age population in NGC~4550. While absorption
linestrengths are not sensitive to the very youngest stars,
they can easily reveal stellar populations 1~Gyr or older.
The age-sensitive linestrength H$\beta$ (measured on the Lick/IDS
system) is relatively
high in NGC~4550, at a value of 1.99 $\rm \AA$ integrated over an
effective radius (R$_{e}$) and
2.14 $\rm \AA$ integrated over R$_{e}$/8 \citep{kuntschner06}, reaching a
value of 2.2 $\rm \AA$
in the very centre (Maier et al., in preparation). Observations with the
Multi-Pupil Fiber Spectrograph (MPFS) measured
lower H$\beta$ linestrengths, however these were not corrected for the
contamination from the H$\beta$ emission line \citep{afanasiev02}. These
linestrength values are hard to reproduce with only old ($> 5$ Gyr)
stars and more likely indicate a young or intermediate age
population, perhaps dating from the time of the merger. As with the
UV, there is again an age-mass fraction 
degeneracy, with a smaller fraction of young stars producing an
analogous effect to a larger fraction of intermediate-age stars.



Unfortunately, neither the UV nor the absorption linestrengths can
conclude anything about ongoing star formation in NGC~4550, while the
FIR only gives an upper limit. To investigate ongoing star formation
further, we look at optical emission line ratios and polyaromatic
hydrocarbon (PAH) emission. Particularly low ratios of  
\oiii to \hbeta emission lines [log([O$\:${\small III}]/H$\beta) < -0.2$] indicate
current star formation ($<20$ Myr) as the source of ionisation
\citep{ho97}. Data from SAURON indicate that log([O$\:${\small
  III}]/H$\beta)$ is between 
0.15 and 0.5 for NGC~4550 \citep{sarzi06}. While gas could be star
forming at these 
high values, it is more probable that shocks, post-asymptotic giant branch stars or an AGN are the 
dominant source of ionisation (Sarzi et al. in preparation). The greater physical extent of the ionised gas
compared to the molecular gas (see Fig.~\ref{fig:ionized}) also supports the idea that star
formation is not the primary ionisation source. Similarly, although the PAH
emission detected in NGC~4550 \citep{bressan06} is suggestive of star
formation, the PAH spectrum lacks the distinctive line ratios seen in
star-forming regions. This different spectrum may just reflect a very low star
formation rate \citep{galliano08}, but it may also indicate another
source of PAH excitation altogether \citep{smith07}. Thus, with the
available data, we cannot say more than that the star formation rate in
NGC~4550 is less than about $0.02$ M$_{\sun}$ yr$^{-1}$, as
indicated by the FIR emission.

At this rate, the amount of molecular gas we detect in NGC~4550 can
fuel star formation for up to 350~Myr. The molecular gas is likely
related to the young population (280~Myr) found in the UV, although whether the
star formation has proceeded continuously or in a more
episodic fashion since then is impossible to say. 
 Depending on the interpretation of the
linestrengths, the star formation period may be of an even longer
duration (extending back to more than $1$~Gyr) or the moderately high
linestrengths may reflect a stellar population originating in the
merger that formed NGC~4550.


\section[]{Conclusions}

We detect a very small amount ($1\times 10^{7}$ M$_{\sun}$) of
molecular gas in the centre of the counter-rotating red disc galaxy
NGC~4550. The CO(1-0) emission is limited to the central 750~pc
(10\arcsec) and is asymmetrically distributed, stronger at positive
relative velocities north of the galaxy centre. The molecular gas
co-rotates with the thick stellar disc and the ionised gas,
counter-rotating with respect to the thin stellar disc.

Our simulation of the merger of two counter-rotating coplanar disc
galaxies shows that the main features of NGC~4550 can naturally be explained with
such a scenario. The interaction heats the prograde disc more than its
retrograde companion and the gas component ends up aligned with the total angular momentum (dominated by the orbital angular momentum), and thus with the prograde disc. Thus, the gas is predicted to always rotate like the dynamically hotter disc, as observed in NGC~4550. As the gas
accretion scenario does not provide a natural explanation for the
corotation of the gas and thick disc, the merger scenario appears the
more likely. 

Both the UV and optical-linestrength data indicate that NGC~4550
cannot be made up of a purely old stellar population. The best-fit
two-population model to the UV-optical photometry in the region with
observed molecular gas gives a young population of 280 Myr and mass of
$5.89 \times 10^7$  $\rm M_{\sun}$. The optical H$\beta$ linestrength
value requires a population at least younger than 5~Gyr. Ongoing star
formation in the observed molecular gas is possible if there are
locally dense regions that exceed the critical density. The FIR
emission gives an upper limit on the current star formation rate of 
$0.02$ M$_{\sun}$ yr$^{-1}$. This low star formation rate combined
with the small amount of molecular gas present suggest that we are
either witnessing a weak period in an extended bursty star formation
episode, or a more continuous very low-level star formation episode
that will last about another 350~Myr.

\section*{Ackowledgments}

The authors acknowledge receipt of a Daiwa Anglo-Japanese Foundation
small grant and a Royal Society International Joint Project grant, 2007/R2-IJP,
which facilitated this work. 
 We would like to thank Philippe Salome
for help with the reduction of the Plateau de Bure data. We are also
grateful to the SAURON Team for providing  SAURON data, including as
yet unpublished MDM and GALEX images. LMY acknowledges
support from grant NSF AST-0507432 and would like to thank the Oxford
Astrophysics Department for its hospitality during sabbatical work. 
MB acknowledges support from NASA through {\it GALEX} Guest Investigator
program GALEXGI04-0000-0109. This work was supported by grant No. R01-2006-000-10716-0 from the
Basic Research Program of the Korea Science and Engineering Foundation
to SKY.

Based on observations carried out with the IRAM Plateau de Bure
Interferometer. IRAM is supported by INSU/CNRS (France), MPG
(Germany) and IGN (Spain). Also based on observations carried out with
the NASA {\it GALEX}. {\it GALEX} is operated for NASA by the
California Institute of Technology under NASA contract NAS5-98034. 
The NASA/IPAC Extragalactic Database (NED) is operated by the Jet
Propulsion Laboratory, California Institute of Technology, under
contract with the National Aeronautics and Space Administration. This
research made use of HyperLEDA: http://leda.univ-lyon1.fr.

\end{document}